\documentclass[]{aastex}
\usepackage{float}
\usepackage{spr-astr-addons}
\usepackage{url}
\usepackage{graphicx}
\usepackage{epstopdf}
\usepackage[fleqn]{amsmath}
\RequirePackage{color}

\usepackage[colorlinks=true,linkcolor=blue,citecolor=blue,urlcolor=blue]{hyperref}%
\usepackage{bm}%

\def\be{\begin{equation}}
\def\ee{\end{equation}}
\def\bea{\begin{eqnarray}}
\def\eea{\end{eqnarray}}
\def\ban{\begin{eqnarray*}}
\def\ean{\end{eqnarray*}}
\def\bd{\begin{displaymath}}
\def\ed{\end{displaymath}}

\def\bc{\begin{center}}
\def\ec{\end{center}}

\def\ba{\begin{array}}
\def\ea{\end{array}}
\usepackage{url}

\usepackage{epsfig}
\usepackage{color}
\usepackage{amsmath}
\usepackage{enumitem}
\usepackage{mathrsfs}
\usepackage{url}
\usepackage[table]{xcolor}
\usepackage{bigints}
\usepackage{soul}
\usepackage{cancel}
\usepackage{multicol}

\begin{document}

\title{Electromagnetic instabilities of low-beta alpha/proton beams in space plasmas}

\slugcomment{Not to appear in Nonlearned J., 45.}
\shorttitle{Electromagnetic ion cyclotron instability}
\shortauthors{Rehman et al.}
\author{M.A.~Rehman$^{1}$ $\bullet$ S.M.~Shaaban$^{2,3\bigstar}$ $\bullet$ P.H.~Yoon$^{4,5,6}$ $\bullet$ M.~Lazar$^{2,7}$ $\bullet$ S.~Poedts$^{2,8}$}

\affil{$^{1}$ Department of Physics, GC University, Kachery Road, Lahore 54000, Pakistan}
\affil{$^{2}$ Centre for Mathematical Plasma Astrophysics, Celestijnenlaan 200B, B-3001 Leuven, Belgium.  \\$^*$ E-mail: \color{blue}{s.m.shaaban88@gmail.com}}
\affil{$^{3}$ Theoretical Physics Research Group, Physics Department, Faculty of Science, Mansoura University, 35516, Egypt.}
\affil{$^{4}$ Institute for Physical Science and Technology, University of Maryland, College Park, MD 20742, USA.}
\affil{$^{5}$ Korea Astronomy and Space Science Institute, Daejeon 34055, Korea;.}
\affil{$^{6}$ School of Space Research, Kyung Hee University, Yongin, Gyeonggi 17104, Korea.}
\affil{$^{7}$ Institut f\"ur Theoretische Physik, Lehrstuhl IV: Weltraum- und Astrophysik, Ruhr-Universit\"at Bochum, D-44780 Bochum, Germany.}
\affil{$^{8}$ Institute of Physics, University of Maria Curie-Sklodowska, 20-400 Lublin, Poland.}

\begin{abstract}
Relative drifts between different species or particle populations are 
characteristic to solar plasma outflows, e.g., in the fast streams of the 
solar winds, coronal mass ejections and interplanetary shocks. This paper 
characterizes the dispersion and stability of the low-beta 
alpha/proton drifts in the absence of any intrinsic thermal anisotropies, 
which are usually invoked in order to stimulate various instabilities. 
The dispersion relations derived here describe the full spectrum of instabilities
and their variations with the angle of propagation and plasma parameters. The 
results unveil a potential competition between instabilities of the 
electromagnetic proton cyclotron and alpha cyclotron modes. For conditions 
specific to a low-beta solar wind, e.g., at low heliocentric distances in 
the outer corona, the instability operates on the alpha cyclotron branch. 
The growth rates of the alpha cyclotron mode are systematically stimulated by the
(parallel) plasma beta and/or the alpha-proton temperature ratio. One
can therefore expect that this instability develops even in the absence
of temperature anisotropies, with potential to contribute to a
self-consistent regulation of the observed drift of alpha particles.  
\end{abstract}

\maketitle

\section{Introduction}\label{sec.1}

Plasma outflows released by the Sun in interplanetary space consist mainly of
electrons, protons and minor ions, among which the alpha particles are dominant
with an average abundance of about 5\% of the total number density of ions.
Owing to their mass density, typically 20\% of the total ion mass density,
alpha-ions may have important implications in the solar wind dynamics
\citep{robbins1970helium, Marsch-2006, kasper2007solar, Maruca2012,
Maneva-etal-2014}. The {\it in-situ} measurements in the fast wind and
regions not too distant from the Sun, i.e., $\lesssim 1$~AU, have revealed
alpha particle streams faster than the protons, with relative speeds of the
order of local Alfv{\'e}n speed \citep{Asbridge1976, Marsch-1982,
neugebauer1981observations}, and kinetic temperatures in general higher
than proton temperature \citep{Kohl1998}. This preferential energization of
heavier ions in collision-poor plasmas is attributed to the interaction with
plasma waves, via Landau or cyclotron resonances \citep{Marsch-2006}, and
is supported by observations \citep{marsch1991kinetic, vonSteiger1995}.

The observed velocity distributions display strong deviations from thermal
equilibrium, often with a dual structure combining a proton core and a
drifting alpha population called alpha beam \citep{Marsch-1982}.
Non-thermal features of the observed ion distributions are sources of free
energy that can locally generate instabilities and enhanced wave fluctuations
\citep{Garybook}. In our case, beaming alpha populations may destabilize the
plasma system via excitations associated with the Alfv\'en/ion cyclotron waves
and the fast magnetosonic/whistler modes \citep{Revathy-1978, Gary-etal-2000a,
Li-Habbal-2000}. In turn, these fluctuations regulate the ion VDFs through
the same wave-particle interactions, which determine a diffusion in velocity
space and a thermalisation and relaxation of the alpha beam
\citep{Marsch-Livi-1987, Gary-etal-2000b}. The observations confirm that
the alphas are heated in perpendicular direction and cool more slowly than
what would be expected from adiabatic expansion \citep{Reisenfeld-etal-2001,
Stansby-etal-2019}.

The investigations of alpha-ion beams have initially concluded that high
plasma beta conditions, e.g., $\beta \geqslant 1$, more typical to large
heliospheric distances ($\geqslant 1$~AU), can be favorable to magnetosonic
instabilities, with a maximum growth for propagation parallel to the
interplanetary magnetic field \citep{Gary-etal-2000a, Gary-etal-2000b}.
Later it was shown that magnetosonic instabilities were stimulated by an
internal anisotropic temperature $T_\perp/T_\parallel < 1$ of protons or
alpha particles ($\parallel, \perp$ being directions with respect to
the magnetic field), while isotropic beams, with, e.g.,
$T_{\alpha, \perp}/T_{\alpha, \parallel} \simeq 1$, may drive another
instability of the parallel-propagating Alfv\'en/ion-cyclotron waves
\citep{Verscharen-etal-2013}. The ion-beam instabilities are also highly
dependent on the plasma beta, which decreases towards the Sun and in
the outer corona ($\sim 0.3$~AU) becomes
$\beta_{p,\alpha}=8\pi n_{p,\alpha}T_{p,\alpha}/B_0^2<1$ for both
the protons (subscript $p$) or alpha particles (subscript $\alpha$)
\citep{Marsch-1982}. Here, $n_{p,\alpha}$ denote proton/alpha number
densities, $T_{p,\alpha}$ their respective temperatures (in the
unit of energy), and $B_0$ denotes the ambient magnetic field intensity.
The low-beta alpha/proton beams appear to be more
susceptible to Alfv\'enic  instabilities, but again, it is not clear
if these instabilities are driven by the alpha/proton drift or by
the anisotropic temperature $T_\perp/T_\parallel > 1$, as such a
condition is always assumed for protons or alpha particles in
the literature \citep{Li-Habbal-2000, Gomberoff-Valdivia-2003}.
In the outer corona an excess of temperature (kinetic energy) in
perpendicular direction may be associated only to the proton core,
while suprathermal populations including proton halo or beams of protons
and alpha particles rather show an opposite, relatively small anisotropy
$T_\perp/T_\parallel \lesssim 1$ \citep{Marsch-1982, Marsch-2006}.

In order to complete the picture and provide a clear basic characterization
of these instabilities, in the present paper we reduce the complexity of
the low-beta beaming plasmas, and investigate alpha/proton drifting
populations with isotropic temperatures $T_\perp/T_\parallel = 1$.
The dispersion and stability properties of the plasma system are derived
on the basis of a general kinetic approach, which cover the full spectrum
of wave-frequencies, wave-numbers and angles of propagation with respect
to the magnetic field. The organization of our paper is the following:
In section~\ref{sec.2}, we derive the general dispersion relation for the
electromagnetic modes propagating at an arbitrary angle. Growth rates of
the unstable solutions are obtained for standard representations of the
particle velocity distribution as drifting-Maxwellians in section \ref{sec.3}.
The unstable solutions are discussed by numerical means in Section~\ref{sec.4},
providing also an analysis of their variation with the main plasma parameters.
Section~\ref{sec.5} summarizes our present results.

\section{Dispersion Relation}\label{sec.2}

In a collision-poor plasma the general linear dispersion relation
for electromagnetic modes propagating at an arbitrary angle $\theta$
with respect to the uniform background magnetic field
($\bm{B}_0=B_0\hat{\bm{z}}$)  is given by
\begin{equation}
0=\det \left|\epsilon_{ij}(\bm{k},\omega)
-\frac{c^2 k^2}{\omega^2}\left(\delta_{ij}
-\frac{k_ik_j}{k^2}\right)\right|,
\label{1}\end{equation}
with a dielectric tensor defined by
\begin{eqnarray}
\epsilon_{ij} &=& \delta_{ij}
+\sum_a\frac{\omega_{pa}^2}{\omega^2}
\int d{\bf v}\left[v_\parallel\left(
\frac{\partial}{\partial v_\parallel}
-\frac{v_\parallel}{v_\perp}
\frac{\partial}{\partial v_\parallel}\right)
b_jf_a\right.
\nonumber\\
&& +\sum_{n=-\infty}^\infty\frac{V_iV_j^*}
{\omega-n\Omega_a-k_\parallel v_\parallel}
\nonumber\\
&& \left.\times\left(\frac{n\Omega_a}{v_\perp}
\frac{\partial}{\partial v_\perp}+k_\parallel
\frac{\partial}{\partial v_\parallel}\right)f_a\right],
\label{HM}\end{eqnarray}
where $\omega$ is the wave frequency, $b_i=\bm{B}_0/B_0=\hat{\bm{ z}}$
is the unit vector along the direction of ambient magnetic field
vector, $f_a(v_\perp, v_\parallel)$ is the velocity distribution
function for particle species labeled $a$ ($a=e,p,\alpha$, etc.,
for electrons, protons, alpha particles, etc.),
$\omega_{pa}=~\sqrt{4 \pi n_a e^2/m_a}$ and
$\Omega_a=e_a B_0/m_a$ are the plasma and cyclotron frequencies
for species $a$, and
\begin{eqnarray}
V_i &=& \left(v_\perp\frac{nJ_n(b)}{b},\mbox{ }
-iv_\perp J'_n(b),\mbox{ }
v_\parallel J_n(b)\right)_i,
\nonumber\\
b &=& \frac{k_\perp v_\perp}{\Omega_a}.
\label{3}\end{eqnarray}
Here, $J_n$ is a Bessel function of the first kind, of order
$n$, with argument $b$. In the above, $e$, $n_a$, $m_a$, and $c$
stand for unit electric charge, ambient density for plasma
particle species $a$, their mass, and the speed of light
{\it in vacuo}, respectively.

In the present analysis we adopt the cold plasma approximation for
the real frequency, combined with warm plasma growth/damping rate
expression in the context of weak growth/damping formula, or equivalent,
the dissipative instability formalism. The weak growth or dissipative
instability theory is also known as the weak kinetic instability
theory, as opposed to the reactive or fluid instability theory. We take
the general cold plasma dielectric tensor elements given
in component form by
\begin{align}
\epsilon_{xx} &=1-\frac{\omega_{pe}^2}{\omega^2-\Omega_e^2}
-\sum_a\frac{\omega_{pa}^2}{\omega^2-\Omega_a^2}=\epsilon_{yy},
\nonumber\\
\epsilon_{xy} &=-i\,\frac{\Omega_e}{\omega}
\frac{\omega_{pe}^2}{\omega^2-\Omega_e^2}
+i\sum_a\frac{\Omega_a}{\omega}
\frac{\omega_{pa}^2}{\omega^2-\Omega_a^2}=-\epsilon_{yx},
\nonumber\\
\epsilon_{xz} &=\epsilon_{zx}=0
=\epsilon_{yz}=\epsilon_{zy},
\nonumber\\
\epsilon_{zz} &=1-\frac{\omega_{pe}^2}{\omega^2}
-\sum_a\frac{\omega_{pa}^2}{\omega^2},
\label{a1}\end{align}
and further approximate the situation by considering two species
ions, namely, protons and alpha particles, and ignore the
displacement current, that is, the unity on the right-hand side of
Eq.\ \eqref{a1}, which is valid for low frequency waves. We also assume
$\omega^2\ll\Omega_e^2$, in order to simplify the dielectric tensor elements,
\begin{align}
\epsilon_{xx} &\approx-\frac{\omega_{pp}^2}{\omega^2-\Omega_p^2}
-\frac{\omega_{p\alpha}^2}{\omega^2-\Omega_\alpha^2},
\nonumber\\
\epsilon_{xy} &=i\,\frac{\omega_{pp}^2}{\omega\Omega_p}
\left(\frac{n_0}{n_p}
+\frac{\Omega_p^2}{\omega^2-\Omega_p^2}\right)
+i\,\frac{\Omega_\alpha}{\omega}
\frac{\omega_{p\alpha}^2}{\omega^2-\Omega_\alpha^2},
\nonumber\\
\epsilon_{zz} &=-\frac{m_p}{m_e}\frac{n_0}{n_p}
\frac{\omega_{pp}^2}{\omega^2},
\label{a5}
\end{align}
where $\Omega_p=eB_0/(m_pc)$, $\Omega_\alpha=\Omega_p/2$,
$\omega_{pp}^2=4\pi n_pe^2/m_p$, $\omega_{pe}^2=m_pn_0\omega_{pp}^2/(m_e n_p)$,  and
$\omega_{p\alpha}^2=(n_\alpha/n_p)\omega_{pp}^2$. Note that in $\epsilon_{zz}$ the other ion terms can be neglected compared to $m_p/m_e\gg1$.

Substituing the dielectric tensor elements \eqref{a5} to the
dispersion relation \eqref{1}, it can be shown that the
linear dispersion relation is given in dimensionless form by
\begin{equation}
q^2=\frac{sx}{(1+\mu^2)x
\mp\sigma\sqrt{(1+\mu^2)^2x^2+st\mu^2}},
\label{a8}\end{equation}
where
\begin{equation}
q=\frac{ck}{\omega_{pp}},\quad
x=\frac{\omega}{\Omega_p},\quad
\mu=\cos\theta,\quad
\delta=\frac{n_\alpha}{n_p},
\end{equation}
are dimensionless wave number, frequency, cosine of the wave
propagation angle, and the alpha-to-proton density ration, and
other quantities are defined by
\begin{eqnarray}
s &=& \frac{2(xQ_-+\delta P_-Q_+)
(xQ_++\delta P_+Q_-)}{D},
\nonumber\\
\sigma &=& \frac{|D|}{D},\quad
D=Q_+Q_-+\delta P_+P_-,
\nonumber\\
t &=& \frac{2P_+P_-Q_+Q_-}{D},
\nonumber\\
P_\pm &=& x\pm1,\qquad Q_\pm=x\pm\tfrac{1}{2}.
\label{a9}\end{eqnarray}

   \begin{figure}[t!]
   \begin{center}
   \includegraphics[width=\columnwidth]{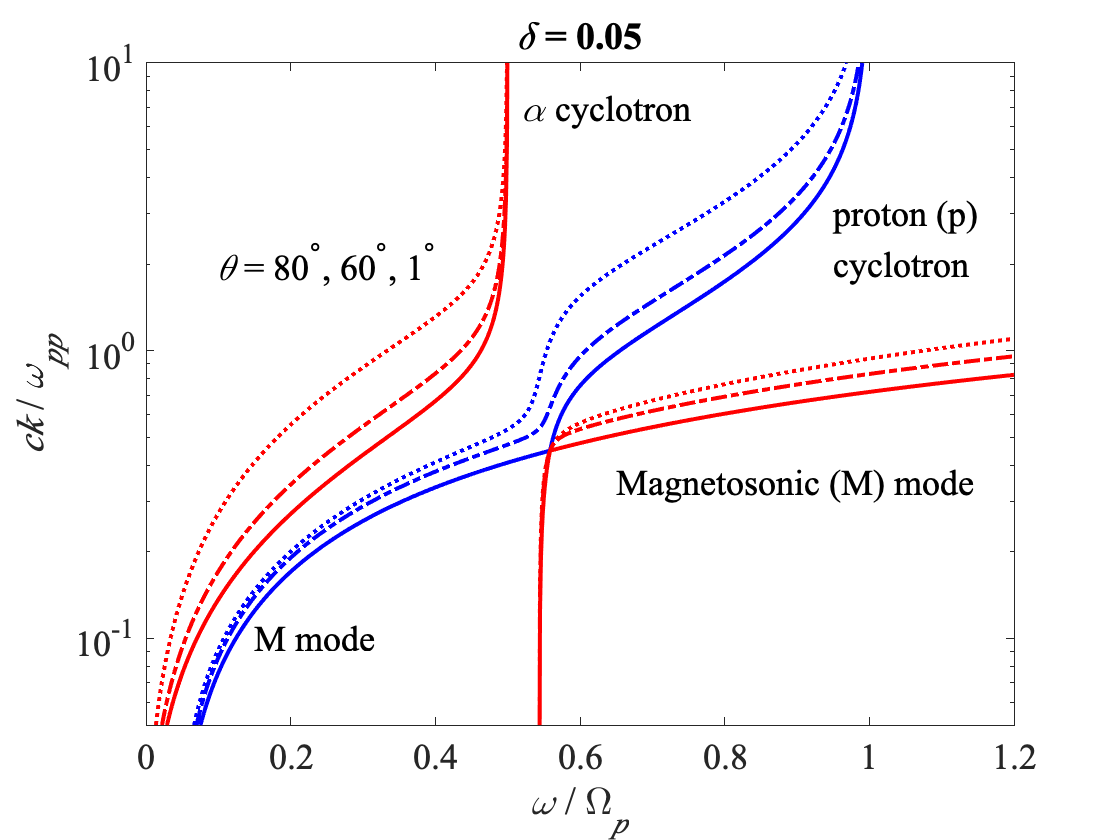}
   \end{center}
   \caption{Dispersion relation plotted as $q^2=c^2k^2/\omega_{pp}^2$
   versus $x=\omega/\Omega_p$ curves for three different values
   of $\theta$ corresponding to $1^\circ$ (solid line), $60^\circ$ (dash-dotted), and
   $\theta=80^\circ$ (dotted).}
   \label{f1}
\end{figure}

Figure \ref{f1} plots $q$ versus $x$ for three values of $\theta$
corresponding to $1^\circ$, $60^\circ$, and $80^\circ$.
The dispersion curves show that they are composed of three
distinct branches. The first branch (topmost, color coded in red)
denotes the low frequency mode that begins as Alfv\'en wave for
low frequency ($\omega\ll\Omega_p$), but gradually turns into
the resonant mode (where $k\to\infty$) at the alpha cyclotron
frequency, $\omega=\Omega_\alpha=\Omega_p/2$.
For quasi parallel propagation angle ($\theta\to0$), this branch
is left-hand circularly polarized.
The second brach (color coded in blue for $\omega<\Omega_p$)
starts off in low frequency regime ($\omega\ll\Omega_p$) as
fast/magnetosonic mode, and this mode is right-hand circularly
polarized for quasi parallel angle of propagation.
For nonzero $\theta$, this mode switches over to the proton
cyclotron mode, which becomes resonant ($k\to\infty$) at the
proton cyclotron frequency ($\omega=\Omega_p$). The switchover
takes place in the vicinity of the alpha cyclotron
frequency. A third brach that begins as the alpha cyclotron
mode turns into the fast/magnetosonic mode above $\Omega_\alpha$
and beyond. The third branch is plotted with red color.
In general, modes designated with red color corresponds to the
lower sign in the dispersion relation \eqref{a8}, while the
blue curves belong to the upper sign.

In the growth rate calculation to be discussed subsequently,
we will denote instabilities operative on the topmost branch as
the ``alpha cyclotron'' instability, while instabilities operative
on the middle, blue-colored mode will be designated as the
``proton cyclotron'' instability. It turns out that the third,
bottommost curve -- that is, the fast/magnetosonic mode --
remains stable.

\section{Growth Rate}\label{sec.3}

Assuming weak growth/damping rate ($|\gamma|\ll\omega$), and following
the standard method of derivation \citep{Melrose}
we obtain an explicit expression for the growth/damping rate,
\begin{eqnarray}\label{eq14}
\gamma &=& \sum_a\frac{\pi\omega_{pa}^2}
{\left[1-|{\bf e}({\bf k})\cdot\hat{\bf k}|^2\right]
\left[\partial(\omega^2N^2)/\partial\omega\right]}
\nonumber\\
&& \times\int d{\bf v}\sum_{n=-\infty}^\infty
\left|{\bf e}({\bf k})\cdot{\bf V}\right|^2
\delta(\omega-n\Omega_a-k_\parallel v_\parallel)
\nonumber\\
&& \times\left(\frac{n\Omega_a}{v_\perp}
\frac{\partial}{\partial v_\perp}+k_\parallel
\frac{\partial}{\partial v_\parallel}\right)
f_a\left(v_\perp, v_\parallel\right),
\end{eqnarray}
where $N=ck/\omega$ is the index of refraction,
${\bf V}$ is defined in Eq.\ \eqref{3}, and ${\bf e}$
represents the unit electric field vector, which is
discussed in Appendix A. Making use of the
dispersion relation \eqref{a8} one may also compute
$\partial(\omega^2N^2)/\partial\omega$ explicitly, which
is given in Appendix B. In Eq.\ \eqref{eq14}
$f_a=f_a(v_\perp^2,v_\parallel)$ represents the velocity
distribution function for ion species labeled $a$.
The following is the resulting explicit expression for the
growth/damping rate after taking into account the
dispersive wave properties associated with the
low frequency modes:
\begin{eqnarray}\label{Rs}
\frac{\gamma}{\Omega_p} &=& \sum_{a=p,\alpha}
\frac{n_a}{n_p}\frac{m_p}{m_a}
\frac{2\pi^2}{(1+M^2x^2\mu^2)R}
\int_0^\infty dv_\perp v_\perp^3
\nonumber\\
&& \times\sum_{n=-\infty}^\infty
\left|M^2x^2\frac{n^2J_n^2}{b^2}
+(J'_n)^2-2Mx\frac{nJ_nJ'_n}{b}\right|
\nonumber\\
&& \times\left(\frac{n\Omega_a}{k v_\perp\mu}
\frac{\partial f_a(v_\perp^2,v_r)}{\partial v_\perp}
+\frac{\partial f_a(v_\perp^2,v_r)}{\partial v_r}\right),
\nonumber\\
v_r &=& \frac{\omega-n\Omega_a}{k\mu},
\end{eqnarray}
where the argument of the Bessel function $b=k_\perp v_\perp/\Omega_a$
applies for proton ($a=p$) and alpha particles ($a=\alpha$),
and $M$ and $R$ are defined in Eqs.\ \eqref{KTM} and \eqref{a48}, respectively.

For drifting Maxwellian distributed plasmas we replace the arbitrary
distribution function $f_a$ in the growth rate expression \eqref{Rs} by
\begin{equation}
f_a(v_\perp,v_\parallel)=\frac{1}{\pi^{3/2}v_{Ta}^3}
\exp\left(-\frac{v_\perp^2}{v_{Ta}^2}
-\frac{(v_\parallel-V_a)^2}{v_{Ta}^2}\right),
\end{equation}
where $v_{Ta}=(2T_a/m_a)^{1/2}$ stands for thermal speed, and
$V_a$ represents the average drift speed.
After some straightforward mathematical manipulations it can be
shown that the growth rate expression reduces to
\begin{eqnarray}\label{e21}
\frac{\gamma}{\Omega_p} &=& -\sum_a\frac{n_a}{n_p}
\frac{m_p}{m_a}\frac{\pi^{1/2}}{(1+M^2\mu^2)R}
\sum_{n=-\infty}^\infty
\nonumber\\
&& \times\biggl(\frac{n(1+M^2x^2)}{2}\,[I_{n-1}(\lambda_a)
-I_{n+1}(\lambda_a)]e^{-\lambda_a}
\nonumber\\
&& +(Mnx+\lambda_a)\,[2I_n(\lambda_a)-I_{n-1}(\lambda_a)
\nonumber\\
&& -I_{n+1}(\lambda_a)]e^{-\lambda_a}\biggr)
\xi_a\exp\left[-(\zeta_n^a)^2\right],
\end{eqnarray}
where $I_n$ is the modified Bessel function of
the first kind of order $n$, and
\begin{align}
&\xi_a=\frac{\omega-kV_a\mu}{kv_{Ta}\mu},
\nonumber\\
& \zeta_n^a=\frac{\omega-n\Omega_a-kV_a\mu}{kv_{Ta}\mu},
\nonumber\\
&\lambda_a =\frac{k^2v_{Ta}^2(1-\mu^2)}{2\Omega_a^2}.
\end{align}

\section{Numerical Analysis}\label{sec.4}

The growth rate of low frequency modes is a function of frequency
$\omega$ and angle of propagation $\theta$. It also depends implicitly
on alpha-proton number density ratio $n_\alpha/n_p$,
alpha particle drift velocity $V_\alpha$ (we assume zero drift
for the protons, $V_p=0$), plasma beta parameters $\beta_p$ and
$\beta_\alpha$, where $\beta_p=8\pi n_pT_p/B_0^2$ and
$\beta_\alpha=8\pi n_\alpha T_\alpha/B_0^2$, respectively.
In the fast solar wind alpha particles possess an average density of 5\%
of the total number density and are drifting with respect to the
protons with a typical speed on the order of local Alfv\'en speed,
$V_\alpha = v_A$ \citep{Marsch-1982, Reisenfeld-etal-2001}.
Consequently, in the present analysis, for alpha-proton relative
number density we consider $n_\alpha/n_p=0.05$, and for alpha drift velocity,
we fix the value at $V_{\alpha}=v_A$. For low corona the beta
values are relatively low. We thus consider low value of $\beta_\alpha=0.01$
and a slightly higher proton beta of $\beta_p=2\beta_\alpha$, as an example.
Note that while $\beta_\alpha$ and $\beta_p$ are comparable, this
actually represent much high alpha particle temperature, since
the alpha particle number density is much lower. This is
consistent with observation.

   \begin{figure*}[t!]
   \begin{center}
      \includegraphics[width=\textwidth]{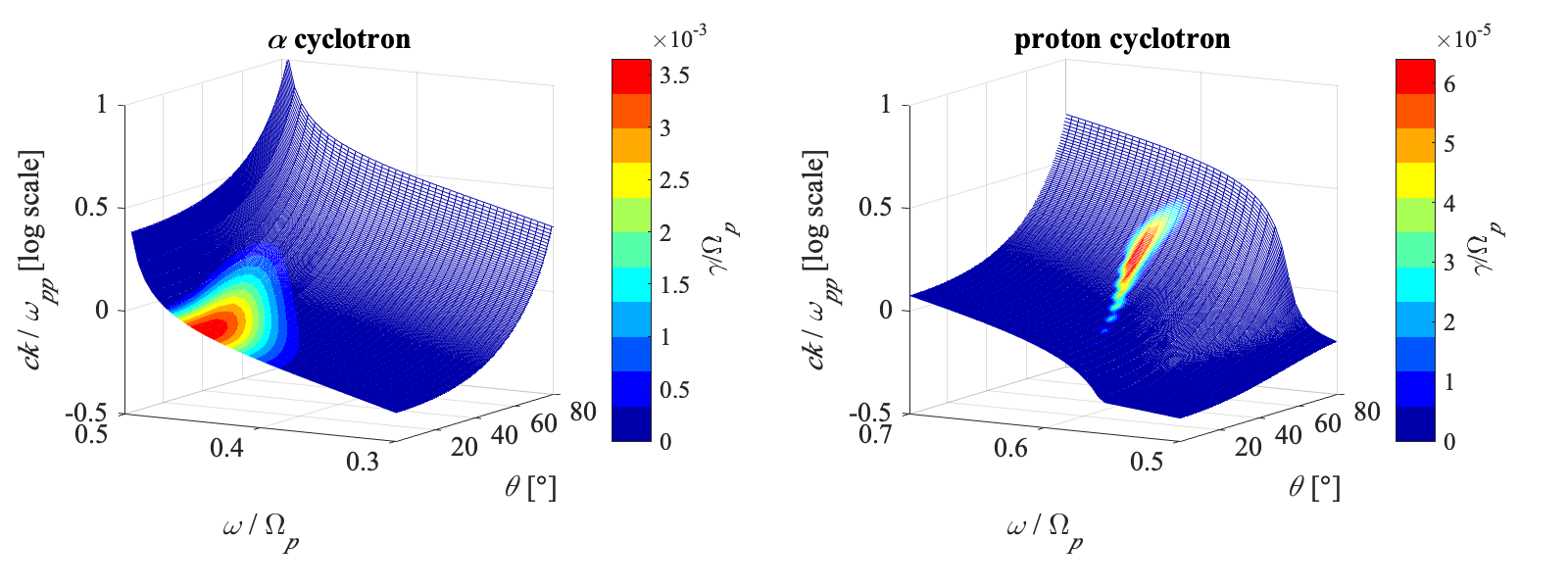}
   \end{center}
   \caption{Dispersion surfaces corresponding to alpha cyclotron
    modes [left], and proton cyclotron mode, which includes
    the magnetosonic mode branch [right]. The color-coded growth
    rates for each mode is shown, for $n_\alpha/n_p=0.05$,
    $T_\alpha/T_p=2$, and $\beta_\alpha=0.01$. Note that the
    alpha cyclotron instability is an order of magnitude higher
    than that for the proton cyclotron instability.}
   \label{f2}
\end{figure*}

Figure \ref{f2} displays on the left, the dispersion surface
or manifold, corresponding to Alfv\'en-alpha cyclotron mode, while
the right-hand panel plots the dispersion surface, depicting the
proton cyclotron branch (which also includes fast/magnetosonic mode
brach in the lower frequency regime).
Vertical axis represents normalized wave number, $q=ck/\omega_{pp}$,
while the two horizontal axes denote normalized (real) frequency
$x=\omega/\Omega_p$, and wave propagation angle $\theta$, respectively.
We indicate the region of wave growth on each surface as well as the
magnitude by color scheme. As indicated by color bars, however, it is
apparent that the alpha cyclotron instability growth rate is almost
an order of magnitude higher than that of proton cyclotron branch.
The instability for both branches take place over narrow bands
of frequencies and along extended domains of propagation angles.
However, only the most unstable alpha cyclotron modes take place along 
quasi parallel direction, while the peaking growth rates of the proton
cyclotron instability peak appear at oblique angles.

\begin{figure}[t!]
   \begin{center}
   \includegraphics[width=\columnwidth]{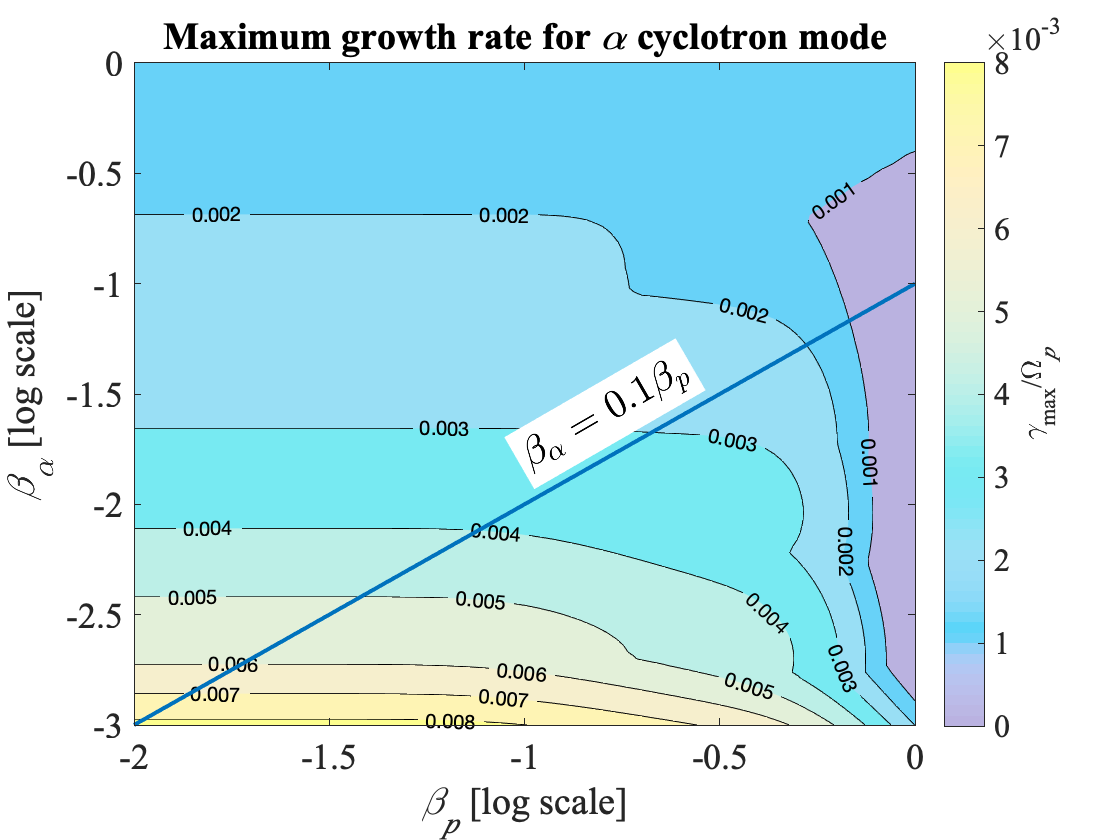}
   \end{center}
   \caption{Maximum growth rate for alpha cyclotron mode branch
   versus $\beta_p$ and $\beta_\alpha$. Other parameters are fixed,
   $n_\alpha/n_p=0.05$ and $V_\alpha=v_A$.} \label{f3}
\end{figure}

In Figure \ref{f3} we plot the maximum growth rate for the unstable
alpha cyclotron modes, which was determined by surveying the entire
frequency and angle space for a given set of input parameters
$\beta_p$ and $\beta_\alpha$. We have then systematically varied
both $\beta_p$ and $\beta_\alpha$, for fixed $\delta=0.05$ and
$V_\alpha=v_A$, until we covered the two dimensional parameter
space $(\beta_p,\beta_\alpha)$.
Figure \ref{f3} shows that the alpha cyclotron beam instability
becomes more unstable as $\beta_\alpha$ decreases, for fixed
value of $\beta_p$. Of course, one cannot indefinitely decrease
$\beta_\alpha$, since the present weak growth/damping rate formalism
assumes that the distribution function has a relatively mild
of velocity space gradient,
$\partial f_\alpha/\partial v_\perp$ and
$\partial f_\alpha/\partial v_\parallel$. For very low $\beta_\alpha$
values, the distribution will have a sufficiently high velocity
derivative so that the assumption of weak growth rate is violated.
Such a caveat notwithstanding, it is interesting to note that
the beam-driven alpha cyclotron instability becomes more unstable
for lower beta values for alpha particles. Figure \ref{f3} shows
that the instability is suppressed as the proton beta (or, equivalently,
proton temperature) increases.

\begin{figure}[t!]
   \begin{center}
   \includegraphics[width=\columnwidth]{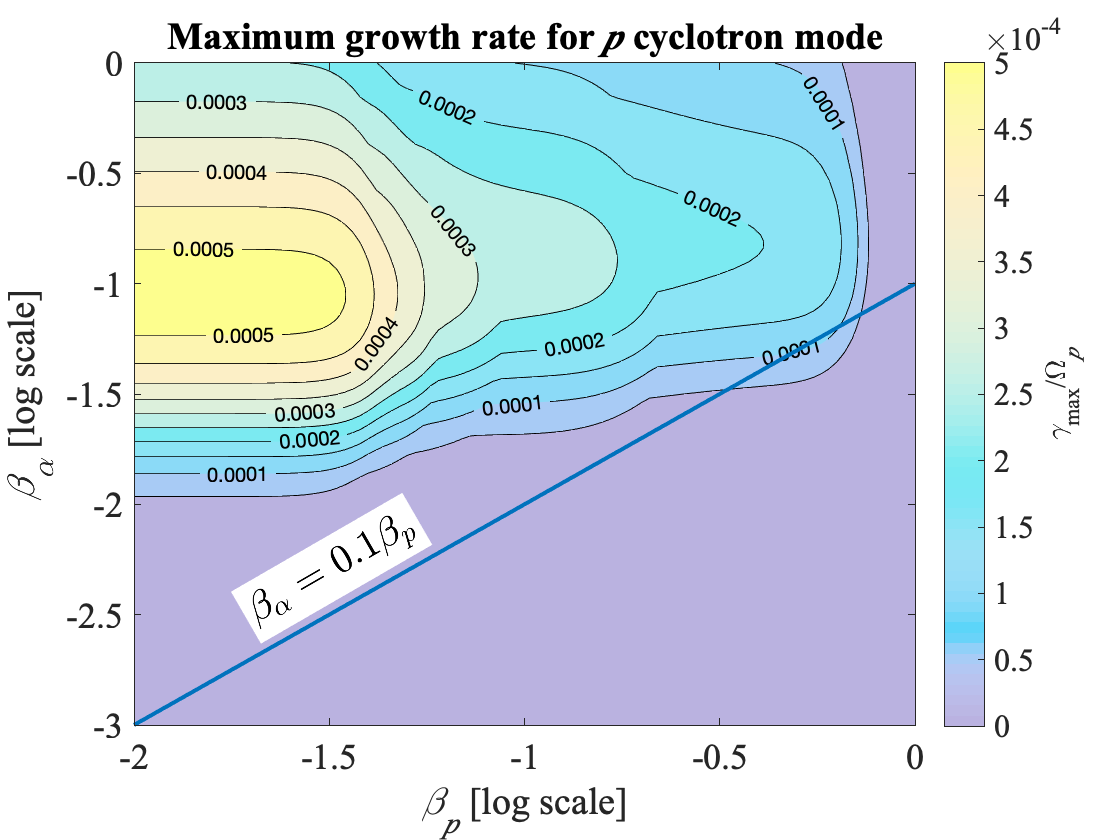}
   \end{center}
   \caption{Maximum growth rate for proton cyclotron mode branch
   versus $\beta_p$ and $\beta_\alpha$. Other parameters are fixed,
   $n_\alpha/n_p=0.05$ and $V_\alpha=v_A$.} \label{f4}
\end{figure}

Figure~\ref{f4} exhibits the maximum growth rate for proton cyclotron
instability in the same format as Fig.\ \ref{f3}. Conditions relevant 
for a low-beta solar wind are in general confined around the solid blue line 
that corresponds to the sample case assumed in Figure~\ref{f2}. Note that the maximum
growth rate is lower in magnitude than that of the alpha cyclotron
instability by an order of magnitude in an overall sense.
It is interesting to note that the maximum proton cyclotron beam
instability has a peak value around $\beta_\alpha\sim10^{-1}$,
but for both higher and lower $\beta_\alpha$,
the maximum growth rate decreases. This behavior is in contrast to that
of the alpha cyclotron instability, where the alpha cyclotron
instability monotonically increases in magnitude of the maximum
growth rate as $\beta_\alpha$ is decreased (until, presumably, the
assumption of weak growth rate is eventually violated).
Note that the proton cyclotron instability is also suppressed by
increasing proton beta, which is similar
to that of the unstable alpha cyclotron mode.

To summarize, we find that the relative proton-alpha beam driven
cyclotron instabilities of both the alpha cyclotron and proton cyclotron
mode branches are generally confined to low beta regime, which in general
conforms with measurements in the solar wind at low altitudes
\citep{Matteini2013, Maruca2012}. Of the two modes, however, the
dominant instability is that of the alpha cyclotron branch, so that
in the nonlinear stage, we expect that the alpha cyclotron mode will
dominate the dynamics. For higher beta values, the beam driven cyclotron
instabilities are generally suppressed, which might explain why in the
literature, the proton-alpha drift instabilities in the high beta regime
have been typically studied in combination with temperature
anisotropies of either the protons or alpha particles.

\section{Conclusions}\label{sec.5}
 
The alpha/proton beam instabilities can play an important role in constraining
the beaming velocity of alpha particles, and may explain the deceleration of
alpha particles in the solar wind with increasing distance from the Sun.
Previous studies have explored in much detail the high beta plasma regime of
these instabilities, clarifying the role of internal temperature anisotropies
of protons or alpha particles, which may switch from Alfv\'enic instabilities
for isotropic beams to an instability of fast-magnetosonic/whistler mode
if alpha beam exhibit $T_\perp / T_\parallel < 1$.

In this paper we have formulated a dissipative (kinetic) theory for the
dispersion and stability of alpha/proton beams, and applied to low
beta regime specific to solar wind conditions at low heliospheric
distances closer to Sun. In this case early studies assume an intrinsic
thermal anisotropy $T_\perp / T_\parallel$ for protons or alpha particles,
which stimulates Alfv\'enic instabilities and prohibits any other
instabilities to develop. Contrary to that, here we assumed alpha and
proton beams with isotropic temperatures $T_\perp/T_\parallel = 1$,
in order to provide basic insights on the dispersion and stability
of alpha/proton beams.

In Section~\ref{sec.2}, we have derived the general dispersion
relations of electromagnetic waves propagating at arbitrary angle 
($\theta$) with respect to the magnetic field. In Section \ref{sec.3},
we formulated the weak growth rate theory for plasma particles that are
assumed to be distributed according to standard (Maxwellian) statistics,
with protons and alpha particles considered to be counter-drifting Maxwellians. 
In section~\ref{sec.4} we have examined a sample growth rate calculation
associated with the electromagnetic modes: fast-magnetosonic/whistler, 
proton-cyclotron and alpha-cyclotron waves, for a given set of alpha-to-proton 
density ratio, $n_\alpha/n_p$, alpha and proton beta's, $\beta_\alpha$ and 
$\beta_p$, and alpha-proton relative drift speed, $V_\alpha$, which is fixed at $v_A$.
The results unveiled a potential competition between instabilities
of proton and alpha cyclotron modes, but the sample calculation also
showed that the alpha cyclotron mode may reach growth rates one order of 
magnitude higher than that of the proton cyclotron mode.
We have then proceeded with the calculation of maximum growth rates for
each mode as we continuously varied $\beta_\alpha$ and $\beta_p$.
It was shown that for the low beta solar wind conditions in the outer corona, 
the alpha cyclotron instability is the dominant mode, with growth rates increasing with
decreasing $\beta_\alpha$. In contrast, the less unstable proton
cyclotron mode has a local peak associated with the maximum growth
rate around $\beta_\alpha=10^{-1}$, but the mode is suppressed for
either decreasing or increasing $\beta_\alpha$ around this peaking value. 
Both modes, however, are stabilized by
increasing proton beta, or equivalently, proton temperature.

The consequence of the excitation of these unstable modes on the
solar wind proton and alpha particle dynamics cannot be understood
purely on the basis of linear theory. In order to address such an
issue, we will carry out quasilinear analysis in the near future.
The impact of instability excitation on the radially expanding
solar wind condition can also be studied in the future where,
the the effects of radial expansion can be counter balanced by
the wave-particle relaxation by instabilities. Such a task
is a subject of our ongoing research.

\begin{acknowledgments}
The authors acknowledge support from the Katholieke Universiteit
Leuven (Grant No. SF/17/007, 2018), 
Ruhr-University Bochum, and Alexander von Humboldt Foundation.
These results were obtained in the framework of the projects SCHL~201/35-1
(DFG--German Research Foundation), GOA/2015-014  (KU Leuven),
G0A2316N (FWO-Vlaa- nderen), and C 90347 (ESA Prodex 9). 
M.A.R. acknowledges Punjab Higher Education Commission (PHEC) Pakistan
for granted Postdoctoral Fellowship FY 2017-18.
S.M.S. gratefully acknowledges support by a Postdoctoral Fellowship
(Grant No. 12Z6218N) of the Research Foundation Flanders (FWO-Belgium).
P.H.Y. acknowledges NASA Grant NNH18ZDA001N-HSR and
NSF Grant 1842643 to the University of Maryland,
and the BK21 plus program from the National Research Foundation
(NRF), Korea, to Kyung Hee University.
\end{acknowledgments}

\appendix

\begin{multicols}{2}
\section{Polarization vector}

For an ambient magnetic field vector directed along $z$ axis,
$\hat{\bf b}={\bf B}_0/|B_0|=\hat{\bf z}$ and the wave vector lying
in $xz$ plane, ${\bf k}=k_\perp\hat{\bf x}+k_\parallel\hat{\bf z}
=\hat{\bf x}k\sin\theta+\hat{\bf z}k\cos\theta$, we define
three orthogonal unit vectors, following \citep{Melrose},
$\hat{\bm\kappa}=\hat{\bf x}\sin\theta+\hat{\bf z}\cos\theta$,
$\hat{\bf a}=\hat{\bf y}$, and
$\hat{\bf t}=\hat{\bf x}\cos\theta-\hat{\bf z}\sin\theta$.
Then the unit wave electric field vector is given by
\begin{equation}
\hat{\bf e}({\bf k})=\frac{\delta{\bf E}}{|\delta E|}
=\frac{K\,\hat{\bm\kappa}+T\,\hat{\bf t}
+i\,\hat{\bf a}}{(K^2+T^2+1)^{1/2}}.
\label{VC}\end{equation}
Making use of linear wave equation,
\begin{align}
\left[\epsilon_{ij}-N^2\left(\delta_{ij}-k_ik_j/k^2\right)\right]
\delta E_j=0,
\end{align}
it is possible to obtain
\begin{align}
\delta E_x &= \frac{\epsilon_{xx}-N^2}{\epsilon_{xy}}\,\delta E_y,
\nonumber\\
\delta E_z &= -\frac{N^2\sin\theta\cos\theta}{\epsilon_{zz}-N^2\sin^2\theta}
\frac{\epsilon_{xx}-N^2}{\epsilon_{xy}}\,\delta E_y.
\label{delta_E}\end{align}
Upon direct comparison with Eq.\ \eqref{VC} one may identify
\begin{align}
K &= -\frac{i\sin\theta\,(\epsilon_{zz}-N^2)\,\epsilon_{xy}}
{\epsilon_{xx}\,\epsilon_{zz}-N^2\,A},
\nonumber\\
T &= -\frac{i\cos\theta\,\epsilon_{zz}\,\epsilon_{xy}}
{\epsilon_{xx}\,\epsilon_{zz}-N^2\,A},
\label{k1t1}\end{align}
where $A = \epsilon_{xx}\sin^2\theta+\epsilon_{zz}\cos^2\theta$.
Upon making use of Eq.\ \eqref{a5}, we further obtain
\begin{align}
K &= -M\sin\theta,\qquad T=-M\cos\theta,
\nonumber\\
M &= \frac{x^2Q_+Q_-+\delta P_+P_-(x^2+\frac{1}{4})}
{x^2D+q^2P_+P_-Q_+Q_-\mu^2},
\label{KTM}\end{align}
where various quantities, $P_\pm$, $Q_\pm$, and $D$,
as well as normalized wave number and frequency,
$q=ck/\omega_{pp}$ and $x=\omega/\Omega_p$, are
defined in Eq.\ \eqref{a9}.

\section{Parameter $R$}

In the growth rate expression \eqref{eq14} appears a quantity
$\partial(\omega^2N^2)/\partial\omega$,
which in normalized form, is defined by
\begin{equation}
R=\frac{\Omega_p}{\omega_{pp}^2}
\frac{\partial(\omega^2N^2)}{\partial\omega}
=\frac{\partial q^2}{\partial x}.
\label{a15}\end{equation}
Making use of Eqs.\ \eqref{a8} and \eqref{a9},
the desired quantity $R$ can readily be computed as
\begin{eqnarray}
R &=& \frac{x}{2P_+P_-Q_+Q_-q^2\mu^2
+D(1+\mu^2)x^2}
\nonumber\\
&& \times\left[\left(\frac{2Q_+Q_-}{P_+P_-}
+\frac{\delta}{2}\frac{P_+P_-}{Q_+Q_-}\right)q^2(1+\mu^2)\right.
\nonumber\\
&& +2\delta^2P_+P_-Q_+Q_-
+\frac{2x^2(x^2-2)Q_+Q_-}{P_+P_-}
\nonumber\\
&& \left.+\frac{\delta x^2}{4}
\frac{8x^4(2x^2-5)+3(9x^2-2)}{P_+P_-Q_+Q_-}\right].
\label{a48}\end{eqnarray}

\bibliographystyle{spr-mp-nameyear-cnd}
\bibliography{References}

\end{multicols}

\end{document}